\documentclass[twocolumn,english,showpacs]{revtex4}
\usepackage{amsmath,amssymb,dcolumn}
\usepackage[dvips]{graphics}
\def\imo{i}
\begin{document}
\title{Quasi-normal modes for black hole solutions unknown in an analytical form}
\author{Alexander Zhidenko}
\email{zhidenko@fma.if.usp.br}
\affiliation{Instituto de F\'{\i}sica, Universidade de S\~{a}o Paulo \\
C.P. 66318, 05315-970, S\~{a}o Paulo-SP, Brazil}
\begin{abstract}
We review the papers \cite{Zhidenko:2005mv,Konoplya:2006rv,Konoplya:2006ar}.
We discuss possibilities of studying the quasi-normal modes of black holes that are not known in an analytical form. Such black holes appear as solutions in various theoretical models and real astrophysical approximations when one takes into account the black hole neighborhood.
\end{abstract}
\maketitle

\section{Introduction}
There are three stages of the perturbation evolution near a black hole. The initial black hole outburst from the perturbation source takes place at early time of the evolution. Then, at late time, the damping (\textit{quasi-normal}) oscillations appear, finally suppressed by power-law or exponential tails at asymptotically late time. The quasi-normal modes (QNMs) can be characterized by a set of complex frequencies: real part is the actual oscillation frequency while imaginary part describes the damping rate of the oscillation \cite{QNMrev}.

The QNMs do not depend on initial perturbations being, thereby, an important characteristic of a black hole. They are expected to be observed in the nearest future with the help of a new generation of gravitational antennas. Therefore, it is important to know how they change when the black hole solution has some corrections due to some matter and fields or due to the possible existence of some physics beyond the classical Einstein gravitation. Mostly, such solutions could be obtained only numerically by resolving some system of differential equations.

According to AdS/CFT (anti-de Sitter/Conformal Field Theory) correspondence, QNMs of a large black hole in AdS space-time should coincide with the poles of the retarded Green function of the perturbation in ${\cal N}=4$ $SU(N)$ super-Yang-Mills theory for large $N$ \cite{Horowitz-Hubeny}. That is why asymptotically AdS black hole solutions are being extensively studied as well.

It turns out that the dominating frequencies depend mostly on the black hole solution behavior in some region near the event horizon. Thus one can find them in frequency domain even if the large distance behavior of the solution is not known at all.

We consider two qualitatively different examples: hairy black hole in anti-de Sitter background and asymptotically flat Einstein-Aether black hole solution. We will show that in both cases the behavior of the solution at large distances is not important because of different reasons:
\begin{itemize}
\item For asymptotically anti-de Sitter background we require Dirichlet boundary conditions at spatial infinity. The most significant part of the metric perturbations stays, thereby, near the black hole. That is why the solution behavior at this region causes dominant influence on the QN spectrum.
\item For the asymptotically flat case the QNMs searching can be reduced to the scattering problem. Being resonant oscillations that survive at late time, the QNMs in this approach are the poles of the reflection coefficient. Therefore, the QN frequencies are determined mainly by the form of the effective potential near its peak.
\end{itemize}

\section{QN spectrum of a black hole in AdS space-time}
Since our approach for the QN modes calculation of asymptotically AdS black hole solutions is based on the Horowitz-Hubeny method \cite{Horowitz-Hubeny} let us start from a brief review of it.

Suppose we have an analytical solution for a spherically symmetric black hole metric. Then any spherically symmetric perturbation after separation of variables $\Phi(t,r)=\exp(-\imo\omega)\phi(r)$ satisfies a second order differential equation which has the form
\begin{equation}\label{radial_part}
\phi''(r)+A(r)\phi'(r)+B(r)\phi(r)=0.
\end{equation}
One should note, that in many cases the equation for the radial part can be reduced to the form (\ref{radial_part}) for the non-spherically symmetric perturbations as well.

By definition, QNMs are eigenvalues $\omega$ of (\ref{radial_part}) with boundary conditions (BCs) which require ingoing wave at the event horizon and zero at spatial infinity \cite{Horowitz-Hubeny}. Such BCs allow us to exclude a regular singularity of the equation (\ref{radial_part}) at the event horizon. After introducing the new variable $z$, that is zero at event horizon and one at spatial infinity, one can obtain the equation
\begin{equation}\label{exp_eq}
s(z)z^2y''(z)+t(z)zy'(z)+u(z)y(z)=0,
\end{equation}
where $y(z)=\phi(z)z^{\imo\omega\kappa}$ and $\kappa>0$ is chosen to make $y(z)$ regular at even horizon ($z=0$).

The Horowitz-Hubeny method works if $s(z)$, $t(z)$ and $u(z)$ are polynoms and all the other singularities of the equation lay outside the unit circle\footnote{This requirement is usually satisfied by the appropriate choice of the variable $z$.}. In this case one can expand $y(z)$ as
\begin{equation}\label{expansion}
y(z)=\sum_{n=0}^{\infty} y_nz^n, \qquad y_0=1.
\end{equation}
The coefficients of the series satisfy the recurrence relation
\begin{equation}\label{yserie}
y_n=-\sum_{k=0}^{n-1}\frac{y_k(k(k-1)s_{n-k}+kt_{n-k}+u_{n-k})}{n(n-1)s_0+nt_0},
\end{equation}
which is finite since $s(z)$, $t(z)$ and $u(z)$ are polynoms:
$$
s(z)=\sum_{n=0}^{M_s}s_nz^n, \quad t(z)=\sum_{n=0}^{M_t}t_nz^n,
\quad u(z)=\sum_{n=1}^{M_u}u_nz^n.
$$

Thus one can find $\omega$ by minimizing $|y(1)|$ restricted by some large number of expansion
terms in (\ref{yserie}).

It is clear that if $s(z)$, $t(z)$ and $u(z)$ are series and $s(1)$, $t(1)$ and $u(1)$ converge quickly enough, then $y_n$ is still possible to calculate because it depends insignificantly on higher terms of the series. In order to find series expansion for $s(z)$, $t(z)$ and $u(z)$, one can use the equations which define the black hole solution. We can always do this because $s(z)$, $t(z)$ and $u(z)$ can be explicitly expressed in terms of the metric coefficients and their derivatives \cite{Zhidenko:2005mv}.

It is important to note that one can control the precision of the eigenvalues $\omega$ by requiring the convergence of the found result with respect to increasing of the number of expansion terms of all the series.

This technique was used in \cite{Zhidenko:2005mv} to study QNMs of the scalar hairy black hole in the AdS background \cite{Winstanley}.

\section{QN spectrum of a black hole in the flat background}
For asymptotically flat black hole solutions, the perturbation equation is usually reduced to a wave-like equation
\begin{equation}\label{wave-like}
\left(\frac{d^2}{dx^2}+\omega^2-V(x)\right)\phi(r)=0.
\end{equation}
The general solution of the equation (\ref{wave-like}) at infinity is
\begin{equation}
\Psi = A_{in} \psi_{in} + A_{out} \psi_{out}, \quad r_{*} \rightarrow \infty.
\end{equation}
The quasi-normal modes in this approach, by the definition, are the poles of the reflection coefficient $A_{out}/A_{in}$. Our
starting point is suggested by the WKB method \cite{WKB} where the asymptotic solutions of the wave equation near the event horizon
and near spatial infinity are matched with Taylor expansion near the peak of the potential, i.e. between the two turning points $V(r) - \omega^{2} = 0$.

\begin{figure}\label{spot}
\caption{Potential for electromagnetic perturbations near the
  Schwarzschild black hole ($r_h=1$, $\ell=2$) and the same potential interpolated
  numerically near its maximum. Even despite the behavior of the two
  potentials are very different in the full region of
  $r$, except for a small region near black hole, low-laying quasi-normal modes
  for both potentials are very close.}
\resizebox{\linewidth}{!}{\includegraphics*{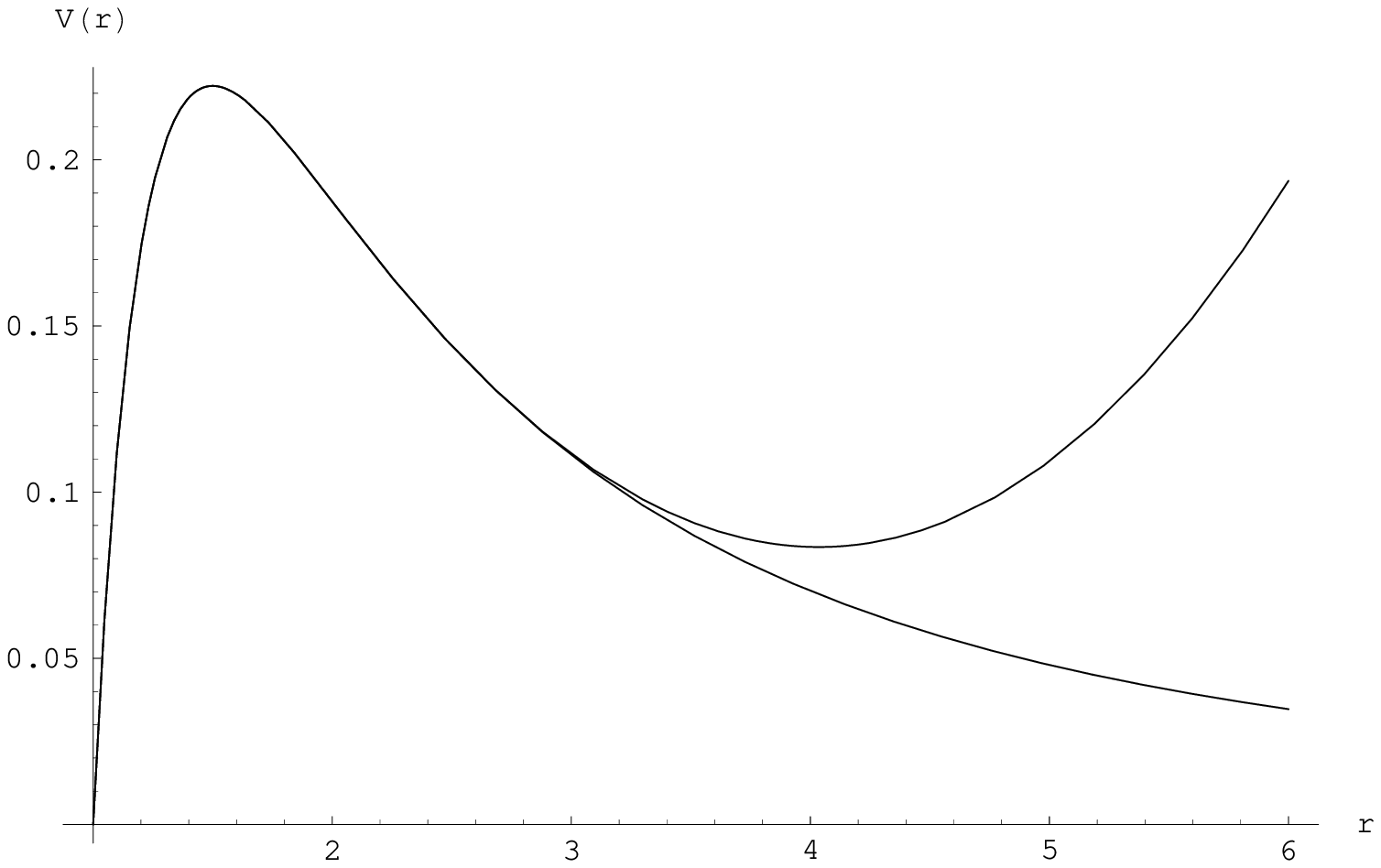}}
\end{figure}
We conclude therefore that the low laying quasi-normal modes are determined mainly by the behavior of the effective potential near
its peak. We shall check that our statement is true by considering the well-known potential for the Schwarzschild black hole $V(x)$ and also two other potentials which lay closely to the Schwarzschild potential near its maximum, but has very different behavior far from a black hole. These two potentials are chosen in the following way. We make a plot of the function $V(x)$ of an analytic Schwarzschild potential, then we find some number of points for this potential $V_{r}$ near its maximum which serve us a basis for our first potential $V_{int}$ which is an {\it interpolation} of these points near the maximum by cubic splines (see Fig. \ref{spot}). Second potential $V_{fit}$ is a {\it fit} of the above plot near the maximum by a ratio of polynomial functions.

The 6-th order WKB formula reads
\begin{equation}\label{WKBformula}
\frac{\imath Q_{0}}{\sqrt{2 Q_{0}''}}
- \sum_{i=2}^{i=6} \Lambda_{i} = n+\frac{1}{2},
\end{equation}
where the correction terms of the i-th WKB order $ \Lambda_{i}$ can be found in \cite{WKBorder}, $Q = V - \omega^2$ and $Q_{0}^{i}$ means the i-th derivative of $Q$ at its maximum. For the WKB formula of sixth order one needs to calculate derivatives of the potential up to 12th order.

Since the WKB formula contains the value of the effective potential its derivatives at the potential peak we find that the results obtained with the help of all three potentials lay very close if the interpolation and the fit were made with the appropriate precision. Despite the higher derivatives of our interpolation potential are not defined, we are able to evaluate them step by step by interpolating the first and all the consequent derivatives of the potential in the same way. Unfortunately, the interpolation potential is very sensitive to numerical errors. Therefore, to calculate the QNMs with the appropriate precision, one must find the values of the potential with very high accuracy. In fact, for the practical purposes one can use fitting of the potential which does not accumulate the numerical error.

To test the accuracy of this approach one can use the convergence of the WKB formula \cite{Konoplya:2006rv}. Since the higher WKB order depends on the higher derivatives of the effective potential, that are more sensitive to the interpolation or fitting accuracy, the higher order WKB formula should give some random values, if the accuracy is not enough.

The possibility of making use of the numerical interpolation or fitting of the potential near its maximum gives us possibility of finding QN spectrum of fields near the solutions which are not known in analytical form. This technique was used in \cite{Konoplya:2006rv,Konoplya:2006ar} to find QNMs of the Einstein-Aether black hole \cite{Jacobson}. The obtained results coincide with those calculated in time domain \cite{Konoplya:2006ar}.

Using this technique we are able to make the next step in studying of the real astrophysical black holes. Such black holes are usually situated in a system of other astrophysical objects (accretion disc, galaxies, dark matter etc.). Any such matter causes influence on the metric and, therefore, changes QN spectrum of the black hole. Since the presented technique does not require neither analytical solution nor its behavior at large distance, it can be used to estimate the influence of the black hole surroundings.

\section{Summary}
To summarize the above information let us briefly remind the main points:
\begin{itemize}
\item The low-laying QNMs depend mostly on the black hole solution behavior in some region near the event horizon.
\item The developed technique allows us to find them if the solution is not known analytically.
\item Within the considered approach we are able to estimate the precision of the obtained results.
\item The approach can be used to find QN spectrum for a wide class of black hole solutions.
\end{itemize}

\begin{acknowledgments}
I would like to thank R.~A.~Konoplya, with whom we developed the method described in \cite{Konoplya:2006rv,Konoplya:2006ar}, for critical reading of this manuscript.\\
This work was supported by \emph{Funda\c{c}\~{a}o de Amparo \`{a} Pesquisa do Estado de S\~{a}o Paulo (FAPESP)}, Brazil.
\end{acknowledgments}


\begin{thebibliography}{99}

%\cite{Zhidenko:2005mv}
\bibitem{Zhidenko:2005mv}
  A.~Zhidenko,
  %``Quasi-normal modes of the scalar hairy black hole,''
  Class.\ Quant.\ Grav.\  {\bf 23} (2006) 3155
  [arXiv:gr-qc/0510039].
  %%CITATION = CQGRD,23,3155;%%

%\cite{Konoplya:2006rv}
\bibitem{Konoplya:2006rv}
  R.~A.~Konoplya and A.~Zhidenko,
  %``Perturbations and quasi-normal modes of black holes in Einstein-aether
  %theory,''
  Phys.\ Lett.\  B {\bf 644} (2007) 186
  [arXiv:gr-qc/0605082].
  %%CITATION = PHLTA,B644,186;%%

%\cite{Konoplya:2006ar}
\bibitem{Konoplya:2006ar}
  R.~A.~Konoplya and A.~Zhidenko,
  %``Gravitational spectrum of black holes in the Einstein-Aether theory,''
  Phys.\ Lett.\  B {\bf 648} (2007) 236
  [arXiv:hep-th/0611226].
  %%CITATION = PHLTA,B648,236;%%

\bibitem{QNMrev} K.~Kokkotas and B.~Schmidt, Living.\ Reviews.\ Relativ. {\bf 2} 2 (1999);\\
H.~P.~Nollert, Class.\ Quant.\ Grav. {\bf 16} R159 (1999).

\bibitem{Horowitz-Hubeny} G.~T.~Horowitz and V.~E.~Hubeny, Phys.\ Rev.\  D {\bf 62} (2000) 024027 [arXiv:hep-th/9909056].

\bibitem{Winstanley}
E.~Winstanley, Found.\ Phys. {\bf 33} (2003) 111-143 [arXiv:gr-qc/0205092]; Class.\ Quant.\ Grav. {\bf 22} (2005) 2233-2248 [arXiv:gr-qc/0501096].

\bibitem{WKB}B.~F.~Schutz and C.~M.~Will Astrophys.\ J.\ Lett {\bf 291} L33 (1985).

\bibitem{WKBorder}S.~Iyer and C.~M.~Will Phys.\ Rev.\  D {\bf 35} 3621 (1987);\\
R. A. Konoplya, Phys.\ Rev\ D {\bf 68}, 024018 (2003).

\bibitem{Jacobson}
  C.~Eling and T.~Jacobson,
  %``Black holes in Einstein-aether theory,''
  Class.\ Quant.\ Grav. {\bf 23} 5643-5660 (2006)
  [arXiv:gr-qc/0604088].

\end{thebibliography}
\end{document}